\documentclass[12pt]{article}
\usepackage{amsmath} 
\input{epsf} 
\def\ltap{\raisebox{-.6ex}{\rlap{$\,\sim\,$}} \raisebox{.4ex}{$\,<\,$}} 
\def\gtap{\raisebox{-.6ex}{\rlap{$\,\sim\,$}} \raisebox{.4ex}{$\,>\,$}}

\newcommand\as{\alpha_{\mathrm{S}}} 
\newcommand\f[2]{\frac{#1}{#2}} 
\def\beq{\begin{equation}} 
\def\eeq{\end{equation}} 
\def\beeq{\begin{eqnarray}} 
\def\eeeq{\end{eqnarray}} 
 
\def\to{\rightarrow}
 
\def\nn{\nonumber}

\def\ms{${\overline {\rm MS}}$}
\def\qt{$q_T$}
\def\bqt{{\bf q_T}}

\def\ep{\epsilon}

\def\qq{$Q{\bar Q}$}
\def\h{{\bf H}}
\def\de{{\bf \Delta}}
\def\v{{\bf V}}
\def\g{{\bf \Gamma}_t}
\def\dcor{{\bf D}}
\def\i{{\bf I}}
\def\L34phi{L_{34}^{\varphi}}

\def\T{{\bf T}}

\def\F{{\bf F}_t^{(1)}}

\def\bpt{{\bf p_T}}
\def\pts{{\bf p}_{\rm \bf T}^2}

\def\c3b{c_{3b}}

\begin{document}
\begin{titlepage}
\renewcommand{\thefootnote}{\fnsymbol{footnote}}
\begin{flushright}
ZU-TH 28/14
\end{flushright}
\vspace*{2cm}

\begin{center}
{\Large \bf Transverse-momentum resummation\\[0.3cm] 
for heavy-quark hadroproduction
}
\end{center}

\par \vspace{2mm}
\begin{center}
{\bf Stefano Catani$^{(a)}$,
Massimiliano Grazzini$^{(b)}$\footnote{On leave of absence from INFN, Sezione di Firenze, Sesto Fiorentino, Florence, Italy.}}
and
{\bf Alessandro Torre$^{(b)}$}

\vspace{5mm}

$^{(a)}$ INFN, Sezione di Firenze and Dipartimento di Fisica e Astronomia,\\ 
Universit\`a di Firenze,
I-50019 Sesto Fiorentino, Florence, Italy\\

$^{(b)}$ Physik-Institut, Universit\"at Z\"urich, 
CH-8057 Z\"urich, Switzerland

\vspace{5mm}

\end{center}

\par \vspace{2mm}
\begin{center} {\large \bf Abstract} \end{center}
\begin{quote}
\pretolerance 10000

We consider the production of a pair of heavy quarks ($Q{\bar Q}$)
in hadronic collisions. When the transverse momentum $q_T$ of the heavy-quark
pair
is much smaller than its invariant mass, the QCD perturbative expansion is
affected by large logarithmic terms that must be resummed to all-orders. This
behavior is well known from the simpler case of hadroproduction of colourless
high-mass systems, such as vector or Higgs boson(s). In the case of $Q{\bar Q}$
production, the final-state heavy quarks carry colour charge and are
responsible for additional soft radiation (through direct emission and
interferences with initial-state radiation)
that complicates the evaluation of the
logarithmically-enhanced terms in the small-$q_T$ 
region.
We present the all-order
resummation structure of the logarithmic contributions,
which includes colour flow evolution factors
due to soft wide-angle radiation. 
Resummation is performed at the completely differential level with 
respect to the kinematical variables of the produced heavy quarks.
Soft-parton radiation produces azimuthal correlations that are fully taken into
account by the resummation formalism.
These azimuthal correlations are entangled with
those that are produced by initial-state collinear radiation.
We present explicit analytical results
up to next-to-leading order and next-to-next-to-leading logarithmic accuracy.

\end{quote}

\vspace*{\fill}
\begin{flushleft}
August 2014

\end{flushleft}
\end{titlepage}

\setcounter{footnote}{1}
\renewcommand{\thefootnote}{\fnsymbol{footnote}}

\section{Introduction}

We consider the inclusive production of a \qq\ pair of heavy quarks ($Q$) in
hadron--hadron collisions. The bulk of the cross section is produced in the
kinematical region where the transverse momentum \qt\ of the \qq\ pair is 
smaller 
than the mass $m$ of the heavy quark. In this paper we are interested in the
small-\qt\ region, namely, the region where $q_T \ll m$ (including the limit
$q_T \to 0$). From the phenomenological point of view, the most relevant 
process is
the production of a pair of top-antitop ($t{\bar t}$) quarks \cite{ttbar},
because of its topical importance in the context of both Standard Model (SM)
and beyond-SM physics. In our theoretical study at the formal level, we consider 
a generic pair of heavy quarks.

The \qt\ cross section of the \qq\ pair is computable in QCD perturbation theory
\cite{fo},
provided $m$ is much larger than the QCD scale $\Lambda_{\rm QCD}$. The cross 
section
is obtained by convoluting the parton densities of the colliding hadrons with 
the partonic cross sections, which are evaluated as power series expansion in 
the QCD coupling $\as$. In the small-\qt\ region the perturbative expansion 
is badly behaved,
since the size of the perturbative coefficients is enhanced by powers of 
$\ln q_T$.
A reliable theoretical calculation requires the all-order resummation of these
logarithmically-enhanced terms. This type of perturbative behaviour is well known
\cite{Dokshitzer:hw, Parisi:1979se, Curci:1979bg}
from the simpler case of hadroproduction of a high-mass lepton pair through 
the Drell--Yan (DY) mechanism. In the case of the DY process the 
all-order resummation 
of the $\ln q_T$ terms is fully understood 
\cite{Dokshitzer:hw, Parisi:1979se, Curci:1979bg, Collins:1984kg}.
At the level of leading-logarithmic (LL) contributions, the extension of
resummation from the DY process to the heavy-quark process is relatively
straightforward, and it was first discussed long ago in Ref.~\cite{Berger:1993yp} (related studies were presented in Ref.~\cite{Mrenna:1996cz}).
Beyond the LL level, the structure of $\ln q_T$ terms for the heavy-quark process
is definitely different (the main physical differences are discussed below)
from that of the DY process, and this difference implies very relevant 
theoretical complications.
The all-order resummation for the heavy-quark process 
has been discussed only very recently by H.~X.~Zhu et al. in 
Refs.~\cite{Zhu:2012ts, Li:2013mia}.
The analysis of Refs.~\cite{Zhu:2012ts, Li:2013mia} is limited to the study 
of the 
\qt\ cross section after integration over the azimuthal angles of the produced 
heavy quarks.
In this paper we illustrate the results of our independent study of  
transverse-momentum resummation for \qq\ production. We present our all-order
resummation formalism for \qq\ production, and we perform the resummation 
up to the next-to-next-to-leading logarithmic (NNLL) level, by explicitly
including 
{\em all} the contributions up to the next-to-leading order (NLO)
in the perturbative expansion. Our formalism and results are valid at the
fully-differential level with respect to the kinematics of the produced heavy 
quarks.
In particular, we consider the explicit dependence on the azimuthal angles 
of the
heavy quarks and we have full control, at the resummed level, of the ensuing 
{\em azimuthal correlations} in the small-\qt\ region.
In the case of the azimuthally-averaged $q_T$ cross section we find agreement with the NNLL results of Refs.~\cite{Zhu:2012ts, Li:2013mia}.

The DY lepton-pair production is a specific process of a general class of
hard-scattering processes in which the produced high-mass system $F$ in the 
final
state is formed by a set of colourless (i.e., non-strongly interacting) particles
(e.g., $F$ can be a lepton pair, or a photon pair, or one or more vector bosons 
or Higgs bosons). Transverse-momentum resummation for the \qt\ distribution 
of $F$ is fully understood for this entire class of processes 
(i.e., independently of the
specific particle content of the system $F$). Indeed, transverse-momentum
resummation for these processes has an all-order {\em universal} 
(process-independent) structure
\cite{Collins:1981uk, Collins:1984kg, Catani:2000vq, Catani:2010pd, 
Catani:2013tia},
which has been explicitly worked out 
\cite{Catani:2013tia, Becher:2010tm, Catani:2011kr, Gehrmann:2012ze}
at NNLL accuracy and the next-to-next-to-leading order (NNLO) in the perturbative
expansion. This universality structure eventually originates from the underlying
physical mechanism that produces the \qt\ broadening of the system $F$ at small 
\qt~: the transverse momentum of $F$ is produced by (soft and collinear) QCD
radiation from the initial-state colliding partons. The \qq\ production process
definitely belongs to a different class of processes, since the produced 
final-state heavy quarks carry colour charge and, therefore, they act as 
additional
source of QCD radiation. The \qt\ of the \qq\ pair depends on initial-state
radiation, on final-state radiation and on quantum (and colour flow) 
interferences
between radiation from the initial and final states. These physical differences
between \qq\ production and the production of a colourless system $F$ lead 
to very relevant technical and conceptual complications in the context of
transverse-momentum resummation for \qq\ production. An important issue regards 
the presence of possible contributions from factorization-breaking effects 
of collinear radiation
\cite{Rogers:2010dm, Catani:2011st, Forshaw:2012bi, Mitov:2012gt}.
Other complications, which already arise in the context of threshold resummation
for the \qq\ total cross section 
\cite{Kidonakis:1997gm, Bonciani:1998vc, Beneke:2009rj, Czakon:2009zw, 
Ahrens:2010zv},
regard the effect of non-abelian colour correlations produced by initial-state 
and final-state interferences. Additional important complications  and effects,
which are specific of transverse-momentum resummation, regard the azimuthal-angle
distribution of the \qq\ pair. In the case of the DY process, \qt\ resummation
has no effect on the azimuthal correlation between the produced leptons, since 
the
\qt\ broadening of the lepton pair is entirely due to QCD radiation from the
initial-state $(q{\bar q})$ partons. In contrast, the \qt\ of the \qq\ pair is 
also due to radiation from $Q$ and ${\bar Q}$ separately, and this leads to 
\qt-dependent azimuthal correlations. The main features of \qq\ production
that we have just highlighted will be briefly recalled in the
presentation of our resummation results.

The paper is organized as follows. In Sect.~2 we introduce our notation and we illustrate our all-order resummation formalism.
In Sect.~3 we present and discuss the explicit form of the resummation coefficients up to NLO and NNLL accuracy.
Our results are summarized in Sect.~4.

\section{All-order resummation}
\label{sec:res}

We consider the inclusive hard-scattering process
\begin{equation}
h_1(P_1)+h_2(P_2)\to Q(p_3)+ {\bar Q}(p_4) + X\, ,
\label{hadpro}
\end{equation}
where the collision of the two hadrons $h_1$ and $h_2$ with momenta 
$P_1$ and $P_2$ produces the \qq\ pair, and $X$ denotes the accompanying
final-state radiation. The hadron momenta $P_1$ and $P_2$ are treated in 
the massless approximation $(P_1^2=P_2^2 \simeq 0)$. The heavy quarks have 
momenta $p_3$ and $p_4$, and the total four-momentum of the \qq\ pair is 
$q^\mu=p_3^\mu+ p_4^\mu$. In a reference frame where the colliding hadrons are
back-to-back, the total momentum $q^\mu$ is fully specified by its invariant 
mass 
$M$ $(M^2=q^2)$, rapidity $y$ $(y =\frac{1}{2} \ln \frac{q\cdot P_2}{q\cdot P_1})$ and
transverse-momentum vector $\bqt$. Analogously, the momentum $p_j^\mu$ $(j=3,4)$
of the heavy quark is specified by the heavy-quark mass $m$ $(p_3^2=p_4^2=m^2)$,
rapidity $y_j$ and transverse-momentum vector ${\bpt}_{j}$.
The two-dimensional transverse-momentum vectors $\bqt$, ${\bpt}_{3}$ and 
${\bpt}_{4}$ have azimuthal angles $\phi_q, \phi_3$ and $\phi_4$.

The kinematics of the observed heavy quarks is fully determined by the their
total momentum $q$ and by two additional and {\em independent} 
kinematical variables
that specify the angular distribution of $Q$ and $\bar Q$ with respect to the
momentum $q$ of the \qq\ pair. These two additional kinematical variables are
generically denoted as ${\bf\Omega}= \{ \Omega_A, \Omega_B\}$
(correspondingly,  we define $d{\bf\Omega} =d\Omega_A \, d\Omega_B$).
For instance, we can use ${\bf\Omega}= \{ y_3, \phi_3 \}$ or any other 
equivalent
pairs of kinematical variables (e.g., $y_3 \to y_3 -y, \;\phi_3 \to \phi_4$ 
and so
forth). We thus consider the most general fully-differential cross section
\begin{equation}
\label{genxs}
\f{d\sigma(P_1, P_2;\bqt,M,y,{\bf\Omega} )}{d^2{\bqt} \;dM^2 \;dy 
\;d{\bf\Omega}} 
\end{equation}
for the inclusive-production process in Eq.~(\ref{hadpro}).
Note that the cross section in Eq.~(\ref{genxs}) and the corresponding 
\qt\ resummation formula can be straightforwardly integrated with respect to 
one or more of the final-state variables $\{\Omega_A,\Omega_B, y, \phi_q, M \}$, thus 
leading to
results for observables that are more inclusive than the differential cross
section in Eq.~(\ref{genxs}).

The hadronic cross section in Eq.~(\ref{genxs}) is computable within QCD by
convoluting partonic cross sections with the scale-dependent parton 
distributions $f_{a/h}(x,\mu^2)$ ($a=q_f,{\bar q}_f,g$ is the label of the
massless partons) of the colliding hadrons. The partonic cross sections are
expressed as a power series expansion in $\as$. At the leading order (LO) in the
perturbative expansion, the partonic cross sections are proportional to $\as^2$
and there are only two contributing partonic processes, namely, the
quark-antiquark ($q{\bar q}$) annihilation process $q_f\,{\bar q}_f \to Q {\bar
Q}$ and the gluon fusion process $g\,g \to Q {\bar Q}$. In both LO processes, the
\qt\ dependence of the partonic cross section (and of the ensuing hadronic cross
section) is simply proportional to $\delta^{(2)}(\bqt)$, because of
transverse-momentum conservation. At higher perturbative orders, the 
partonic cross sections receive contributions from elastic 
($c{\bar c} \to Q {\bar Q}$) and inelastic ($a\,b \to Q {\bar Q} + X$)
partonic processes. The \qt\ dependence of the partonic cross section 
includes contributions that are `singular' in the limit $q_T \to 0$:
these singular contributions are proportional to 
$\as^{n+2} \delta^{(2)}(\bqt)$ or to logarithmic terms of the type 
$\as^{n+2} \frac{1}{q_T^2} \ln^k(M^2/q_T^2)$ with $k \leq 2n -1$
(more precisely, the logarithmic terms are expressed in terms of singular, though
integrable over \qt, `plus'-distributions). We thus decompose the cross section
in Eq.~(\ref{genxs}) as follows:
\begin{equation}
\label{sigdec}
d\sigma = d\sigma^{({\rm sing})} + d\sigma^{({\rm reg})} \;\;,
\end{equation}
where  the component $d\sigma^{({\rm sing})}$ embodies {\em all} the singular
terms in the limit $q_T \to 0$, whereas $d\sigma^{({\rm reg})}$ includes the
remaining non-singular terms. In this paper we deal with the all-order evaluation
and resummation of the small-\qt\ singular terms in $d\sigma^{({\rm sing})}$.
At fixed value of \qt\, the cross section depends on the mass scales $M$ and $m$.
We use $M$ to set the scale of the $\ln q_T$ terms, and the remaining 
dependence on the two mass scales is controlled by the dimensionless ratio
$2m/M$ or, equivalently, by the relative velocity $v$ of $Q$ and ${\bar Q}$,
\begin{equation}
\label{relv}
v= \sqrt{1 - \frac{m^4}{(p_3\cdot p_4)^2}} = 
\sqrt{1 - \left(\frac{2m^2}{M^2-2m^2}\right)^2} \;\;.
\end{equation}
In our resummation treatment at small \qt, the mass scales 
$M$ and $m$ are considered to be
parametrically of the same order. In two particular regions, namely, the
threshold region where $2m/M \to 1$ (or $v \to 0$) and the high-mass region
where $2m/M \to 0$ (or $v \to 1$), the size of the coefficients of the 
$\ln q_T$ terms can be enhanced, and accurate quantitative predictions may
require additional resummation of the dependence on $2m/M$ (or $v$).
Note, however, that our treatment of the small-\qt\ dependence is valid in the
{\em entire} region $q_T \ll M$ (and not only in the subregion $q_T \ll m$).
In other words, in our treatment of the small-\qt\ region, the decomposition in
Eq.~(\ref{sigdec}) is such that we have
$d\sigma^{({\rm reg})}/d\sigma^{({\rm sing})} = {\cal O}(q_T/M)$ 
(modulo logarithmic corrections) order-by-order
in the perturbative QCD expansion (note that 
${\cal O}(q_T/M) \ll {\cal O}(q_T/m)$ if $m \ll M$).

Our discussion of the decomposition in Eq.~(\ref{sigdec}) can be expressed in a
more formal way. We consider the order-by-order perturbative expansion of the
$q_T$ cross section $d\sigma$ and we write $d\sigma = \sum_n d\sigma^{(n)}$,
where $d\sigma^{(n)}$ is the contribution at the $n$-th perturbative order in
$\as$. Analogous perturbative expansions apply to $d\sigma^{({\rm sing})}$ and
$d\sigma^{({\rm reg})}$ in terms of the $n$-th order contributions
$d\sigma^{({\rm sing}) (n)}$ and $d\sigma^{({\rm reg}) (n)}$, and we have
$d\sigma^{(n)} = d\sigma^{({\rm sing}) (n)} + d\sigma^{({\rm reg}) (n)}$.
The regular component of the $q_T$ cross section is thus specified by requiring
that the integration of $d\sigma^{({\rm reg})}/d^2 \bqt$ over the range 
$0 \leq  q_T \leq Q_0$ leads to a finite result that, at each fixed order in
$\as$, {\em vanishes} in the limit $Q_0 \to 0$. We have
\begin{equation}
\int_0^{Q_0^2} dq_T^2 \;\f{d\sigma^{({\rm reg}) (n)}}{d^2{\bqt} \;dM^2 \;dy 
\;d{\bf\Omega}} = {\cal O}\left(Q_0/M\right) \;\;, \quad Q_0 \to 0 \;,
\nn
\end{equation}
and we note that the right-hand side is power suppressed through the ratio
$Q_0/M$.  This requirement on $d\sigma^{({\rm reg})}$ uniquely specifies all the
singular terms of $d\sigma$ that are included in $d\sigma^{({\rm sing})}$,
although there is still some freedom on how non-singular terms (i.e., terms
leading to corrections of ${\cal O}(Q_0/M)$ in the limit $Q_0 \to 0$)
are split between 
$d\sigma^{({\rm reg}) (n)}$ and $d\sigma^{({\rm sing}) (n)}$.
In the following we present an explicit all-order expression of 
$d\sigma^{({\rm sing})}$ (see Eq.~(\ref{crosssec})). This expression can
systematically be expanded in powers of $\as$ thus leading to the explicit
expression of $d\sigma^{({\rm sing}) (n)}$. The explicit
expression of $d\sigma^{({\rm sing}) (n)}$ then uniquely determines 
$d\sigma^{({\rm reg}) (n)}$ in terms of the complete perturbative expression of
the $q_T$ cross section (i.e., we have $d\sigma^{({\rm reg}) (n)}
= d\sigma^{(n)} - d\sigma^{({\rm sing}) (n)}$). More detailed discussions on the 
decomposition in Eq.~(\ref{sigdec}) and on its perturbative expansion can be
found in Refs.~\cite{Catani:2010pd,Bozzi:2005wk}.

We illustrate the method that we have used to derive
our resummation results for $d\sigma^{({\rm sing})}$.
More details and additional results will be presented in forthcoming studies.
We carry out our analysis of
the singular terms in the small-\qt\ region by working in impact parameter 
$(\bf b)$ space and, thus, we first perform the Fourier transformation of 
$d\sigma^{({\rm sing})}/d^2 \bqt$ with respect to $\bqt$ at fixed $\bf b$.
The final results for $d\sigma^{({\rm sing})}/d^2 \bqt$ are then eventually
recovered by performing the inverse Fourier transformation from 
$\bf b$ space to $\bqt$ space (see Eq.~(\ref{crosssec})). 
In $\bf b$ space the singular
terms are proportional to power of $\ln (Mb)$ ($q_T \ll M$ corresponds to
$bM \gg 1$). These $\ln (Mb)$  terms are produced by the radiation of soft and
collinear partons (i.e., partons with low transverse momentum $k_T$, say, with
$k_T \ll M$) in the inclusive final state $X$ of the inelastic 
partonic processes $a\,b \to Q {\bar Q} + X$. Soft and collinear radiation is
treated by using the universal (process-independent) all-order factorization
formulae
\cite{Kosower:1999xi, Bern:1999ry, Catani:1999ss, Catani:2000pi,
Catani:2003vu, Catani:2011st, Feige:2013zla}
of QCD scattering amplitudes. Soft/collinear factorization at the amplitude
(and squared amplitude) level is not spoiled by kinematical effects at the cross
section level, since we are working in  $\bf b$ space
(in the small-\qt\ limit, the kinematics of the \qt\ cross section is exactly
factorized \cite{Parisi:1979se} 
by the Fourier transformation to $\bf b$ space).
Therefore, the $\ln (Mb)$  terms are explicitly computed by the phase space
integration (in $\bf b$ space) of the soft/collinear factors. The application of
the known explicit expressions
\cite{Campbell:1997hg, Bern:1999ry, Kosower:1999rx, Catani:1999ss, 
Catani:2000pi, Czakon:2011ve}
of soft/collinear factorization formulae allows us to compute the structure of
$d\sigma^{({\rm sing})}$ up to NNLO and NNLL accuracy.
The method that we have just described is completely analogous 
(as applied in the NNLL+NLO computation of Ref.~\cite{deFlorian:2000pr}
and outlined to all orders in Ref.~\cite{Catani:2013tia})
to the method that is applicable to transverse-momentum resummation for the
production of a system $F$ of colourless particles. The differences between the
production of $F$ and $Q{\bar Q}$ production are due to the non-abelian colour
charge of the produced heavy quarks.
The complications that arise from these differences are basically related to
soft radiation at wide angles with respect to the direction of the colliding
partons. As a consequence, the structure of $d\sigma^{({\rm sing})}$
for $Q{\bar Q}$ production definitely {\em differs} 
(and the differences already appear at the NLO) 
from that of 
transverse-momentum resummation for the production of a colourless system 
$F$. Beyond the NNLL+NNLO level of perturbative accuracy,
non-abelian soft wide-angle interactions of absorptive origin
produce violation of strict factorization for space-like collinear radiation
\cite{Catani:2011st}.
Therefore, the all-order formula of $d\sigma^{({\rm sing})}$ that is presented
below is based on some assumptions about possible contributions that can arise 
from factorization-breaking effects of collinear radiation
\cite{Rogers:2010dm, Catani:2011st, Forshaw:2012bi, Mitov:2012gt}.
In particular, we assume that infrared divergences produced by inclusive parton
radiation at transverse momentum $k_T \ll 1/b$ are either cancelled or
customarily factorized in the parton distributions $f_{a/h}(x,1/b^2)$
evolved up to the scale $\mu \sim 1/b$. Moreover, our resummed result for 
$d\sigma^{({\rm sing})}$ includes only the possible soft/collinear correlation
structures that we have explicitly uncovered up to NNLL+NNLO.
These issues certainly deserve further and future investigations.
We remark that we have full control of the all-order structure of 
$d\sigma^{({\rm sing})}$ up to NNLL+NNLO accuracy. Possible additional
structures are likely to be absent till very high perturbative orders
\cite{Rogers:2010dm, Catani:2011st, Forshaw:2012bi, Forshaw:2006fk}.

In the following we use parton densities $f_{a/h}(x,\mu^2)$ as defined
in the \ms\ factorization scheme.
The running coupling $\as(\mu^2)$ denotes the renormalized QCD coupling in the
\ms\ renormalization scheme with decoupling of the heavy quark $Q$
\cite{Bernreuther:1981sg}
(e.g., in the case of $t {\bar t}$ production,  $\as(\mu^2)$ is the 
\ms\ coupling in the 5-flavour scheme), and $m$ is the renormalized pole mass
of the heavy quark $Q$. Obviously our explicit results can be straightforwardly
expressed in different factorization/renormalization schemes by applying the
corresponding scheme transformation relations (e.g., the pole mass $m$ can be
replaced by the  \ms\ running mass $m(\mu^2)$).
To present the resummation results for $Q{\bar Q}$ production we closely follow
the formulation of transverse-momentum resummation for the production 
of a colourless system $F$, and we use the same notation as in 
Refs.~\cite{Catani:2010pd, Catani:2013tia}
(more details about the notation can be found therein).
This presentation allows us to clearly identify and highlight the structural
differences that arise in the context of $Q{\bar Q}$ production.

Our results for the singular component $d\sigma^{({\rm sing})}$ of the 
$Q{\bar Q}$ production cross section are given by the following all-order
resummation formula:

\begin{align}
\label{crosssec}
&\f{d\sigma^{({\rm sing})}(P_1, P_2;\bqt,M,y,{\bf\Omega} )}{d^2{\bqt} \;dM^2 \;dy \;d{\bf\Omega}} 
=\f{M^2}{2P_1\cdot P_2}\sum_{c=q,{\bar q},g} \;
\left[d\sigma_{c{\bar c}}^{(0)}\right]
\int \f{d^2{\bf b}}{(2\pi)^2} \;\, e^{i {\bf b}\cdot \bqt} \;
  S_c(M,b)\nn \\
& \;\;\;\; \times \;
\sum_{a_1,a_2} \,
\int_{x_1}^1 \f{dz_1}{z_1} \,\int_{x_2}^1 \f{dz_2}{z_2} 
\; \left[ \left( \h \,\de \right) C_1 C_2 \right]_{c{\bar c};a_1a_2}
\;f_{a_1/h_1}(x_1/z_1,b_0^2/b^2)
\;f_{a_2/h_2}(x_2/z_2,b_0^2/b^2) \;
\;, 
\end{align}
where $b_0=2e^{-\gamma_E}$
($\gamma_E=0.5772\dots$ is the Euler number) is a numerical coefficient,
and the kinematical variables $x_1$ and $x_2$ are 
\begin{equation}
\label{xifract}
x_1= \f{M}{\sqrt{2P_1\cdot P_2}} \;e^{+y} \;\;, \quad \quad
x_2=\f{M}{\sqrt{2P_1\cdot P_2}} \;e^{-y} \;\;.
\end{equation}
The right-hand side of Eq.~(\ref{crosssec}) involves the (inverse)
Fourier transformation
with respect to the
impact parameter ${\bf b}$ and two convolutions over the 
longitudinal-momentum fractions $z_1$ and $z_2$.
The parton densities 
$f_{a_i/h_i}(x,\mu^2)$ of the colliding hadrons are evaluated at the scale
$\mu= b_0/b$, which depends on the impact parameter.
The factor that is denoted by the symbol
$\left[d\sigma_{c{\bar c}}^{(0)}\right]$ refers to the {\em partonic}
elastic-production process $c{\bar c} \to Q{\bar Q}$ of the \qq\ pair,
\begin{equation}
c(p_1)+{\bar c}(p_2)\to Q(p_3)+ {\bar Q}(p_4) \;\;,
\quad \quad c=q,{\bar q},g \;\;,
\label{partpro}
\end{equation}
with
\begin{equation}
p_i= x_i P_i \;\;, \quad i=1,2 \;\;,
\end{equation}
where $P_i$ $(i=1,2)$ are the momenta of the colliding hadrons 
(see Eq.~(\ref{hadpro})) and $x_i$ $(i=1,2)$ are the momentum fractions in 
Eq.~(\ref{xifract}). Making the symbolic notation explicit, the symbol 
$\left[d\sigma_{c{\bar c}}^{(0)}\right]$ is related the LO cross section
$d{\hat \sigma}^{(0)}$ for \qq\ production by the partonic process in 
Eq.~(\ref{partpro}), and we have
\beq
\left[d\sigma_{c{\bar c}}^{(0)}\right] = \as^2(M^2) \,
\frac{d{\hat \sigma}_{c{\bar c}\to Q {\bar Q}}^{(0)}(p_1,p_2;p_3,p_4)
}{M^2 \,d{\bf\Omega}} \;\;.
\eeq

QCD radiative correction are embodied in the factors $S_c$ and
$\left[ \left( \h \,\de \right) C_1 C_2 \right]$ on the right-hand side of
Eq.~(\ref{crosssec}). The expression in Eq.~(\ref{crosssec}) involves the sum
of two types of contributions, which correspond to the LO partonic channels:
the contribution of the $q{\bar q}$ annihilation channel $(c=q,{\bar q})$ and
the contribution of the gluon fusion channel $(c=g)$.
In each of these channels, the structure of Eq.~(\ref{crosssec}) is {\em
apparently} similar to the structure of transverse-momentum resummation for the
production of a colourless system $F$ 
\cite{Collins:1981uk, Collins:1984kg, Catani:2000vq, Catani:2010pd,
Catani:2013tia}
(see Eq.~(6) of 
Ref.~\cite{Catani:2013tia} for direct comparison).
The important {\em differences} that occur in the case of \qq\ production 
are hidden in the symbolic notation of the factor 
$\left[ \left( \h \,\de \right) C_1 C_2 \right]$ and, more specifically, they
are due to the factor $\de$ that is related to the accompanying {\em soft}-parton
radiation in \qq\ production. In the case of production of a 
colourless system $F$ , the factor $\de$ is absent (i.e. $\de =1$).

The expression of the symbolic factor 
$\left[ \left( \h \,\de \right) C_1 C_2 \right]$ for the 
$q{\bar q}$ annihilation channel is
\begin{align}
\label{what}
\left[ \left( \h \,\de \right) C_1 C_2 \right]_{c{\bar c};a_1a_2}
 = \left( \h \,\de \right)_{c{\bar c}}
\;\, C_{c \,a_1}(z_1;\as(b_0^2/b^2)) 
\;\, C_{{\bar c} \,a_2}(z_2;\as(b_0^2/b^2)) \;\;, \quad (c=q,{\bar q})\;\;,
\end{align}
whereas for the gluon fusion channel $(c=g)$ we have
\beq
\label{whatgg}
\left[ \left( \h \,\de \right) C_1 C_2 \right]_{gg;a_1a_2}
\!= \left( \h \,\de \right)_{gg;\mu_1 \,\nu_1, \mu_2 \,\nu_2 }
\, C_{g \,a_1}^{\mu_1 \,\nu_1}(z_1;p_1,p_2,{\bf b};\as(b_0^2/b^2)) 
\, C_{g \,a_2}^{\mu_2 \,\nu_2}(z_2;p_1,p_2,{\bf b};\as(b_0^2/b^2)) 
\,. 
\eeq
The functions $C_{ca}$ and $C_{ga}^{\mu \nu}$ are described below.
The factors $\left( \h \,\de \right)$ in Eqs.~(\ref{what}) and (\ref{whatgg})
depend on $\bf b$, $M$ and on the kinematical variables of the partonic process
in Eq.~(\ref{partpro}) (this dependence is not explicitly denoted in 
Eqs.~(\ref{what}) and (\ref{whatgg})). Equation (\ref{whatgg}) includes the sum
over the repeated indices $\{ \mu_i, \nu_i \}$, which refer to the Lorentz
indices of the colliding gluons $g(p_i) \;(i=1,2)$ in Eq.~(\ref{partpro}).
In Eqs.~(\ref{what}) and (\ref{whatgg}) we use the
shorthand notation
$\left( \h \,\de \right)$ for the contribution of
the factors $\h$ and $\de$, since these factors embody
a non-trivial dependence on the colour structure (and colour indices) of the
partonic process in Eq.~(\ref{partpro}). To take into account the colour
dependence, we use the colour space formalism of 
Ref.~\cite{Catani:1996vz}:
the colour-index dependence of the scattering amplitude
${\cal M}$ of the process in Eq.~(\ref{partpro}) is represented by a vector
$| \,{\cal M} \, \rangle$ in colour space, and colour matrices are represented
by colour operators acting onto $| \,{\cal M} \, \rangle$. Using the 
colour space formalism, we can write the explicit representation of 
$\left( \h \,\de \right)$. In the case of the 
$q{\bar q}$ annihilation channel, we have
\begin{equation}
\label{Hq}
\left( \h \,\de \right)_{c{\bar c}}
=\f{\langle \,\widetilde{\cal M}_{c{\bar c}\to Q {\bar Q}} \,|
\; \de \; | \,\widetilde{\cal M}_{c{\bar c}\to Q {\bar Q}} \, \rangle
}{\as^{2}(M^2)
\;|{\cal M}_{c{\bar c}\to Q {\bar Q}}^{(0)}(p_1, p_2;p_3,p_4)|^2}\;\; ,
\quad (c=q,{\bar q})\;\;,
\end{equation}
where the `hard-virtual' amplitude 
$\widetilde{\cal M}_{c{\bar c}\to Q {\bar Q}}$ is directly related to the
infrared-finite part of the all-order (virtual) scattering amplitude
${\cal M}_{c{\bar c}\to Q {\bar Q}}$ of the partonic process in 
Eq.~(\ref{partpro}), and ${\cal M}_{c{\bar c}\to Q {\bar Q}}^{(0)}$ is the
tree-level (LO) contribution to ${\cal M}_{c{\bar c}\to Q {\bar Q}}$
( $|{\cal M}_{c{\bar c}\to Q {\bar Q}}^{(0)}|^2$ is the squared amplitude
summed over the colours and spins of the partons $c,{\bar c},Q,{\bar Q}$).
The relation between ${\cal M}$ and $\widetilde{\cal M}$ is given in 
Eq.~(\ref{mtil}).
The analogue of Eq.~(\ref{Hq}) in the gluon fusion channel is
\begin{equation}
\label{Hg}
\left( \h \,\de \right)_{gg;\mu_1 \,\nu_1, \mu_2 \,\nu_2 }
=\f{\langle \,\widetilde{\cal M}_{gg 
\to Q {\bar Q}}^{\nu_1^\prime \nu_2^\prime} \,|
\; \de \; | \,\widetilde{\cal M}_{gg 
\to Q {\bar Q}}^{\mu_1^\prime \mu_2^\prime} \, \rangle
\;d_{\mu_1^\prime \mu_1} \;d_{\nu_1^\prime \nu_1}
\;d_{\mu_2^\prime \mu_2} \;
 d_{\nu_2^\prime \nu_2}
}{\as^{2}(M^2)
\;|{\cal M}_{gg\to Q {\bar Q}}^{(0)}(p_1, p_2;p_3,p_4)|^2}\;\; ,
\end{equation}
where $\{ \mu_i^\prime, \nu_i^\prime \}$ $(i=1,2)$ are exactly 
(see Eq.~(\ref{mtil})) the gluon Lorentz
indices of the scattering amplitude
${\cal M}_{gg\to Q {\bar Q}}(p_1, p_2;p_3,p_4)$, and 
$d^{\,\mu \nu}= d^{\,\mu \nu}(p_1,p_2)$ is the following
polarization tensor, 
\begin{equation}
\label{dten}
d^{\,\mu \nu}(p_1,p_2) = - \,g^{\mu \nu} + 
\f{p_1^\mu p_2^\nu+ p_2^\mu p_1^\nu}{p_1 \cdot p_2} \;\;,
\end{equation}
which projects onto the Lorentz indices in the transverse plane.
The soft-parton factor $\de$ depends on colour matrices, and it acts as a 
colour space
operator in Eqs.~(\ref{Hq}) and (\ref{Hg}). We can also introduce 
a colour space operator $\h$ through the definition 
$\as^2 \,|{\cal M}^{(0)}|^2 \,\h = 
| \widetilde{\cal M} \, \rangle \,
\langle \,\widetilde{\cal M} |$. Therefore, according to Eqs.~(\ref{Hq}) 
and (\ref{Hg}), the shorthand
notation $\left( \h \,\de \right)$ is equivalent to
$\left( \h \,\de \right) = {\rm Tr} \left[ \h \,\de \right]$, where `${\rm Tr}$'
exactly denotes the colour space trace of the colour operator $\h \,\de$.

We now illustrate the structural form of the resummation formulae in 
Eqs.~(\ref{crosssec}), (\ref{what})--(\ref{Hg}), and the differences between 
\qq\
production and the production of a colourless system $F$. The hard factor $\h$
is independent of the impact parameter ${\bf b}$, and it depends on the
scattering amplitude ${\cal M}_{c{\bar c}\to Q {\bar Q}}$. An analogous 
process-dependent hard factor (which depends on the scattering amplitude of the
process $c{\bar c} \to F$)
\cite{Catani:2013tia}
appears for the production of a colourless system $F$. The functions 
$C_{c \,a}$ \cite{Catani:2000vq}
and $C_{g\,a}^{\mu\,\nu}$ \cite{Catani:2010pd}
in Eqs.~(\ref{what}) and (\ref{whatgg}) are {\em universal} (they are process
independent and only depend on the parton indices), and they are computable as
power series expansions in $\as(b_0^2/b^2)$.
These functions originate from initial-state collinear radiation of partons with
typical transverse momentum $k_T \sim 1/b$.
The function $S_c(M,b)$ in Eq.~(\ref{crosssec}) is the Sudakov form factor
\cite{Collins:1981uk},
and it is also {\em universal}. Thus, for instance, the $q{\bar q}$ annihilation
channel functions $S_q$ and $C_{q \,a}$ also contribute to transverse-momentum
resummation for the DY process
\cite{Collins:1984kg},
whereas the gluon fusion channel functions $S_g$ and $C_{g\,a}^{\mu\,\nu}$
also contribute in the case of Higgs boson production
\cite{Catani:2010pd}. The Sudakov form factor $S_c(M,b)$ resums logarithmic
terms $\as^n \ln^k(Mb)$, starting from the LL contributions (those with $k=2n$)
to the $q_T$ cross section.
The Sudakov form factor is due to QCD radiation from the initial-state partons
$c$ and $\bar c$ in the process of Eq.~(\ref{partpro}) and, more precisely,
it is produced by soft and flavour-conserving collinear radiation 
with typical transverse momentum 
$k_T$ in the range $1/b \ltap k_T \ltap M$. The factor $\de$ in 
Eqs.~(\ref{crosssec}), (\ref{what})--(\ref{Hg}) is specific of \qq\ production
($\de=1$ for the production of a colourless system $F$), and it is due to
QCD radiation of soft non-collinear (at wide angles with respect to the
direction of the initial-state partons) partons from the underlying subprocess 
$c{\bar c}\to Q {\bar Q}$. Therefore, $\de$ embodies the effect of soft
radiation from the \qq\ final state and from initial-state and final-state
interferences. As in the case of the Sudakov form factor, the soft radiation
contribution to $\de$ involves the transverse-momentum  
range $1/b \ltap k_T \ltap M$. Therefore, $\de$ resums additional logarithmic
terms $\as^n \ln^k(Mb)$ (see Eq.~(\ref{deeq})), although the dominant 
contributions
to $\de$ are of next-to-leading-logarithmic (NLL) type, since they are produced
by non-collinear radiation. Moreover, soft-parton radiation at the scale 
$k_T \sim 1/b$ has a `special' physical role, since it is eventually 
responsible for azimuthal correlations (see Eqs.~(\ref{deeq}) and (\ref{deq})).

The soft-parton factor $\de$ depends on the impact parameter $\bf b$, 
on $M$ and on
the kinematics of the partonic process in Eq.~(\ref{partpro}). To explicitly
denote the kinematical dependence (which is in turn related to the two angular
variables $\bf \Omega$ of the $q_T$ cross section), we use the rapidity
difference $y_{34}=y_3 -y_4$ between $Q(p_3)$ and ${\bar Q}(p_4)$ and the
azimuthal angle $\phi_3$ of the quark $Q(p_3)$ (the dependence on $2m/M$ is not
explicitly denoted in the following). The all-order structure of $\de$ is
\beq
\label{deeq}
\de({\bf b}, M;y_{34},\phi_3)  = \v^{\dagger}(b, M;y_{34}) \; 
\;\dcor(\as(b_0^2/b^2);\phi_{3b}, y_{34})
\;\v(b, M;y_{34}) \;\;,
\eeq
where
\beq
\label{veq}
\v(b, M;y_{34}) = {\overline P}_q \;
\exp \left\{ - \int_{b_0^2/b^2}^{M^2} \frac{dq^2}{q^2} \; \g(\as(q^2);y_{34}) 
 \right\} 
\;\;,
\eeq
\beq
\label{gaeq}
\g(\as;y_{34}) = \frac{\as}{\pi} \;\g^{(1)}(y_{34}) + 
\left(\frac{\as}{\pi}\right)^{\!2} \;\g^{(2)}(y_{34}) +
\sum_{n=3}^\infty \left(\frac{\as}{\pi}\right)^n \;\g^{(n)}(y_{34}) \;\;,
\eeq
\beq
\label{deq}
\dcor(\as;\phi_{3b}, y_{34}) = 1 + \frac{\as}{\pi}\;
\dcor^{(1)}(\phi_{3b}, y_{34}) +
\sum_{n=2}^\infty \left(\frac{\as}{\pi}\right)^n 
\;\dcor^{(n)}(\phi_{3b}, y_{34}) \;\;.
\eeq
The colour operator (matrix) $\g$ 
is the soft anomalous
dimension matrix that is specific of transverse-momentum resummation for \qq\
production. This quantity is computable order-by-order in $\as$ as in 
Eq.~(\ref{gaeq}). The {\em evolution} factor $\v$ in Eq.~(\ref{veq}) is obtained
by the exponentiation of the integral of the soft anomalous
dimension. The integral is performed over the transverse-momentum scale 
$q^2$ of the the QCD running coupling, and the symbol ${\overline P}_q$ in
Eq.~(\ref{veq}) denotes the anti path-ordering of the exponential matrix with
respect to the integration variable $q^2$. The evolution operator $\v$
explicitly resums logarithmic terms $\as^n(M^2) \ln^k(Mb)$ (with $k \leq n$)
through
the integration over $q^2$. Soft-parton radiation from the process $c{\bar c}
\to Q{\bar Q}$ produces non-abelian colour correlations that are embodied in
the soft anomalous dimension matrix. The structure of $\v$ is {\em typical}
of the resummation of soft-gluon logarithmic contributions in QCD multiparton
hard-scattering processes 
\cite{Kidonakis:1998nf, Bonciani:2003nt}.
Operators that are analogous to $\v$ arise in the context of threshold
resummation for the \qq\ total cross section
\cite{Kidonakis:1997gm, Bonciani:1998vc, Beneke:2009rj, Czakon:2009zw, 
Ahrens:2010zv}.
The colour operator $\dcor$ in Eq.~(\ref{deeq}) is computable as a powers series
expansion in $\as(b_0^2/b^2)$ (see Eq.~(\ref{deq})). This operator does not
explicitly depend on the hard scale $M^2$, and its dependence on the scale
$b^2$ is due to the running coupling $\as(b_0^2/b^2)$. Therefore, the operator
$\dcor$ effectively resums $\ln(Mb)$ contributions to $\de$ by using the
renormalization group evolution of $\as(\mu^2)$ to express $\as(b_0^2/b^2)$
in terms of $\as(M^2)$ and $\ln(M^2b^2)$. 

An important point about the structure of the soft factor $\de$ in 
Eq.~(\ref{deeq}) regards its dependence on the rapidity and azimuth kinematical
variables of the \qq\ pair. Both $\g$ and $\dcor$ depend on $y_{34}$ and this
produces an ensuing dependence of the operators $\v$ and $\de$.
The azimuthal dependence is {\em specific} of transverse-momentum resummation.
In particular, we remark that $\g$ and, thus, the evolution operator $\v$ do not
depend on azimuthal angles. In contrast, the operator $\dcor$ does depend on
$\phi_3$ and, more importantly, it depends on $\phi_{3b} = \phi_3 - \phi_b$,
where $\phi_b$ is the azimuth of the two-dimensional impact parameter vector
$\bf b$. Inserting this dependence on $\phi_{3b}$ in the resummation formula
(\ref{crosssec}) and performing the inverse Fourier transformation from 
$\bf b$ space to $\bqt$ space, we obtain an ensuing dependence of the
$\bqt$ cross section on $\phi_3 - \phi_q$ (where $\phi_q$ is the azimuthal 
angle of $\bqt$). In other words, the resummation formula (\ref{crosssec})
leads to \qt-dependent {\em azimuthal correlations} of the produced \qq\ pair
in the small-\qt\ region. These azimuthal correlations are produced by the
dynamics of soft-parton radiation, and they are entirely embodied in the
soft-parton factor $\dcor$ of Eq.~(\ref{deeq}). The $\phi_{3b}$ dependence
occurs in $\dcor$, at the characteristic scale $1/b$, and does not occur
in the evolution
operator $\v$: this fact has a definite physical origin in the distinction
between real and virtual radiative contributions. Virtual radiation involves
soft partons with transverse momentum $k_T$ in the entire range $k_T \ltap M$,
while real radiation is due to partons with $k_T \ltap q_T \sim 1/b$.
The dynamics of $\v$ is essentially driven by soft {\em virtual} partons,
which cannot produce azimuthal correlations. Real radiation plays a `minimal' 
role in $\v$: it simply produces the cancellation of virtual terms (and the
ensuing infrared divergences) in the region $k_T \ltap 1/b$, thus leading to
remaining contributions from the region $1/b \ltap  k_T \ltap M$
(see the limit of integrations over $q \sim k_T$ in Eq.~(\ref{veq})).
Azimuthal correlations are instead necessarily produced by {\em real} radiation,
which first occur at scale $k_T \sim q_T \sim 1/b$: these correlations are thus
`trapped' in the soft factor $\dcor(\as(b_0^2/b^2))$, at the corresponding scale
$1/b$.

As first pointed out in Ref.~\cite{Catani:2000vq}, the structure of
transverse-momentum resummation is invariant under a class of renormalization
group transformations, named resummation-scheme transformations. 
This symmetry permits a redefinition of the individual
resummation factors in such a way that their total contribution to the \qt\
cross section is left unchanged. In particular, we can consider a 
resummation-scheme transformation that changes (redefines) the separate factors 
$\h, \v$ and $\dcor$ in such a way that 
$\left( \h \,\de \right)$ (i.e., Eqs.~(\ref{Hq}) and (\ref{Hg})) is invariant.
Such a transformation can introduce an arbitrary $\phi_{3b}$ dependence of the
redefined factors $\h, \v, \dcor$. Our key point about the structure of the
azimuthal correlations in Eq.~(\ref{deeq}) is that there are necessarily schemes
in which the dependence on $\phi_b$ is absent from $\h$ and $\v$, and it is
entirely embodied in $\dcor$. This key point eventually follows from our
previous discussion on the physical origin of the soft-parton azimuthal
correlations. In particular, we can define the factor $\dcor$ in 
Eq.~(\ref{deeq}) in such a way that it gives a trivial contribution after
azimuthal average over $\bf b$. Thus, the soft factor $\dcor$ can fulfils the
property
\begin{equation}
\label{dav}
\langle \;\dcor(\as;\phi_{3b}, y_{34}) \rangle_{{\rm av.}} =1 \;\;,
\end{equation}
where the symbol $\langle \dots \rangle_{\rm av.}$ denotes the azimuthal average
over the angle $\phi_b$ of the impact parameter vector $\bf b$.

We note that the transverse-momentum resummation formula (\ref{crosssec})
has an additional source of azimuthal correlations. These additional
azimuthal correlations are due to the $\bf b$ dependence
of the function $C_{g \, a}^{\mu \, \nu}$ that contributes to 
Eq.~(\ref{whatgg}). The two sources of azimuthal correlations have a definitely
different physical origin. The azimuthal correlations produced by 
$C_{g \, a}^{\mu \, \nu}$ originate from initial-state collinear radiation
\cite{Catani:2010pd}, while those produced by $\dcor$ originate from soft
radiation in the processes, such as \qq\ production, with final-state coloured
partons. This difference is manifest in the $q{\bar q}$ annihilation channel, 
where we
find soft-parton azimuthal correlations (produced by $\dcor$) without
accompanying azimuthal correlations of collinear origin (see Eq.~(\ref{what})).

The gluon collinear function $C_{g \, a}^{\mu \, \nu}$ of Eq.~(\ref{whatgg})
has the following all-order form \cite{Catani:2010pd}:
\begin{equation}
\label{cggten}
C_{g \,a}^{\,\mu \nu}(z;p_1,p_2,{\bf b};\as) =
d^{\,\mu \nu}(p_1,p_2) \;C_{g \,a}(z;\as) + D^{\,\mu \,\nu}(p_1,p_2;{\bf b}) 
\;G_{g \,a}(z;\as) \;\;,
\end{equation}
where $d^{\,\mu \nu}$ is given in Eq.~(\ref{dten}),
\begin{equation}
\label{dbten}
D^{\,\mu \nu}(p_1,p_2;{\bf b}) = d^{\,\mu \nu}(p_1,p_2) - 
2 \; \f{b^\mu \,b^\nu}{\bf b^2} \;\;,
\end{equation}
and $b^\mu = (0,{\bf b},0)$ is the two-dimensional impact parameter vector
in the four-dimensional notation $(b^\mu b_\mu = - {\bf b^2})$.
The perturbative expansion of $C_{g \,a}$ ($a=q,{\bar q},g$)
starts at ${\cal O}(1)$
($C_{g \,a}(z;\as) = \delta(1-z) \,\delta_{g \,a} + {\cal O}(\as)$),
analogously to the collinear functions $C_{q \,a}$ and $C_{{\bar q} \,a}$
in Eq.~(\ref{what}), whereas the expansion of the gluonic function 
$G_{g \,a}$ starts at ${\cal O}(\as)$.
From Eq.~(\ref{cggten}) we see that the dependence of $C_{g \,a}^{\,\mu \nu}$
on the azimuthal angle $\phi_b$ of ${\bf b}$ is entirely embodied in the Lorentz
tensor $D^{\,\mu \nu}$ of Eq.~(\ref{dbten}): therefore, this azimuthal dependence
is uniquely 
specified at arbitrary perturbative orders in $\as$.
This specific azimuthal dependence is a consequence \cite{Catani:2010pd}
of the fact that gluonic collinear radiation is 
intrinsically 
spin-polarized and its spin-polarization structure is uniquely specified 
(see, e.g., Eq.~(50) in Ref.~\cite{Catani:2010pd}) by helicity conservation rules.
The contribution of the gluon fusion channel is the sole source of azimuthal
correlations 
\cite{Nadolsky:2007ba, Catani:2010pd} in transverse-momentum
resummation for the production of a colorless system $F$.
The azimuthal dependence of $C_{g \,a}^{\,\mu \nu}$ produces a definite structure
of azimuthal correlations with respect to the azimuthal angle $\phi_q$ of
the transverse momentum $\bqt$. As shown in Ref.~\cite{Catani:2010pd}, the
small-$\bqt$ resummed cross section for the production of a colourless 
system $F$ through gluon fusion leads to azimuthal correlations that are expressed
in terms of a linear combination of {\em only} four Fourier harmonics
($\cos (2\phi_q), \sin (2\phi_q),\cos (4\phi_q), \sin (4\phi_q)$).

In the case of \qt\ resummation for \qq\ production, the azimuthal dependence is
present in both the $q{\bar q}$ annihilation channel and the gluon fusion channel.
In both channels, the $\phi_b$ dependence of the resummation formula
(\ref{crosssec}) is embodied in the resummation factors at scale $b_0^2/b^2$,
which are (see Eqs.~(\ref{what}), (\ref{whatgg}) and (\ref{deeq}))
\begin{align}
\label{phiq}
&\dcor \;C_{c \,a_1} \;C_{{\bar c} \,a_2}   \;\;\;\;  \quad(c=q,{\bar q}) \;, \\
\label{phig}
&\dcor \;C_{g \,a_1}^{\,\mu_1 \nu_1} \;C_{g \,a_2}^{\,\mu_2 \nu_2}   \;\;,
\end{align}
where we have omitted the argument of the various factors to shorten the notation.
As we have just recalled, the azimuthal dependence of the collinear function
$C_{g \,a}^{\,\mu \nu}$ is relatively simple and it is uniquely specified to all
perturbative orders. In contrast, the $\phi_b$ dependence of $\dcor$ is determined
by the process-dependent dynamics of soft-parton radiation in \qq\ production:
this dependence is definitely cumbersome already at the first perturbative order
(see Eq.~(\ref{d1cor})),
and it receives additional contributions to each subsequent order. Therefore, the
ensuing azimuthal correlations of the \qt\ cross section depend on Fourier
harmonics of any degrees. In particular, in the gluon fusion channel 
(see Eq.~(\ref{phig})), the azimuthal dependence originating from soft-parton
radiation is entangled with the azimuthal dependence of collinear origin:
the complete azimuthal dependence is determined by a non-trivial interplay of
colour (soft) and spin (collinear) correlations.

The resummation formula (\ref{crosssec}) can be straightforwardly averaged over
the azimuth $\phi_q$ of $\bqt$. The resummation formula for the 
azimuthally-averaged \qt\ cross section is obtained from Eq.~(\ref{crosssec})
through two simple replacements: the integrand factor $e^{i {\bf b}\cdot \bqt}$
is replaced by the $0$-th order Bessel function $J_0(bq_T)$ and the factors in
Eq.~(\ref{phiq}) and (\ref{phig}) are replaced by their azimuthal average over
$\phi_b$. Performing the azimuthal average over $\phi_b$, we have
\begin{align}
\label{aveq}
&\langle \; \dcor \;C_{c \,a_1} \;C_{{\bar c} \,a_2} \rangle_{{\rm av.}} 
=  C_{c \,a_1} \;C_{{\bar c} \,a_2} \;\;\;\;  \quad(c=q,{\bar q}) \;, \\
\label{aveg}
&\langle \; \dcor \;C_{g \,a_1}^{\,\mu_1 \nu_1} \;C_{g \,a_2}^{\,\mu_2 \nu_2}  
\rangle_{{\rm av.}} \neq  
\langle \;C_{g \,a_1}^{\,\mu_1 \nu_1} \;C_{g \,a_2}^{\,\mu_2 \nu_2}
\rangle_{{\rm av.}}\;\;.
\end{align}
Owing to the property in Eq.~(\ref{dav}), the effect of the soft-parton factor
$\dcor$ disappears from the right-hand side of Eq.~(\ref{aveq}): therefore, 
in the 
$q{\bar q}$ annihilation channel, soft wide-angle radiation contributes to the 
azimuthally-averaged \qt\ cross section only through the evolution factor
$\v^\dagger \v$ from Eq.~(\ref{deeq}).
Despite the property in Eq.~(\ref{dav}), however, in the gluon fusion channel
we have the inequality in Eq.~(\ref{aveg})
(owing to Eq.~(\ref{dav}) and the fact that $G_{g \,a}={\cal O}(\as)$,
the inequality is due to contributions at ${\cal O}(\as^2)$).
Therefore, the soft factor $\dcor$
still gives a non-trivial effect to the 
azimuthally-averaged \qt\ cross section through the contribution of the gluon
fusion channel. This effect is proportional to the factor 
$\langle \; \dcor \;C_{g \,a_1}^{\,\mu_1 \nu_1} \;C_{g \,a_2}^{\,\mu_2 \nu_2}  
\rangle_{{\rm av.}}$, which originates from the entangled soft/collinear
azimuthal dependence of the \qt\ resummation formula (\ref{crosssec}).

The contribution of the hard factor $\h$ to Eqs.~(\ref{Hq}) and (\ref{Hg}) is
independent of $\bf b$, it depends on the hard scale $M$ and it is entirely
specified by the hard-virtual amplitude
$\widetilde{\cal M}_{c{\bar c}\to Q {\bar Q}}$. The auxiliary amplitude 
$\widetilde{\cal M}_{c{\bar c}\to Q {\bar Q}}$ is related to the scattering
amplitude ${\cal M}_{c{\bar c}\to Q {\bar Q}}$ by the following all-order
factorization formula:
\beq
\label{mtil}
| \,\widetilde{\cal M}_{c{\bar c}\to Q {\bar Q}}(p_1, p_2;p_3,p_4)  \rangle
= \left[ 1 - 
\widetilde{\i}_{c{\bar c}\to Q {\bar Q}}(\as(M^2), \ep)
\right]
\;| {\cal M}_{c{\bar c}\to Q {\bar Q}}(p_1, p_2;p_3,p_4)  \rangle \;\;,
\eeq
where
\begin{align}
\widetilde{\i}_{c{\bar c}\to Q {\bar Q}}(\as(M^2), \ep) =
\frac{\as(\mu_R^2)}{2\pi} 
\;\widetilde{\i}^{(1)}_{c{\bar c}\to Q {\bar Q}}(\ep,M^2/\mu_R^2)
+ \sum_{n=2}^{\infty} 
\left(\frac{\as(\mu_R^2)}{2\pi} \right)^{\!\!n}
\;\widetilde{\i}^{(n)}_{c{\bar c}\to Q {\bar Q}}(\ep,M^2/\mu_R^2)
\;\;,
\label{itilall}
\end{align}
and $\mu_R$ is the renormalization scale. The function 
$\widetilde{\i}_{c{\bar c}\to Q {\bar Q}}(\as, \ep)$ also depends on the momenta
$p_i$ ($i \leq 4$), although this dependence is not explicitly denoted in its
argument.
The structure of Eq.~(\ref{mtil}) is analogous \cite{Catani:2013tia}
to that of the hard-virtual
amplitudes of transverse-momentum resummation for the production of colourless
systems $F$. The main technical difference regards the colour treatment and,
thus, the `subtraction' operator $\widetilde{\i}_{c{\bar c}\to Q {\bar Q}}$ is a
colour operator acting onto the colour vector  
$| {\cal M}_{c{\bar c}\to Q {\bar Q}} \rangle$.

The all-order (virtual) amplitude of the process $c{\bar c}\to Q {\bar Q}$ has
ultraviolet (UV) and infrared (IR) divergences. We consider their regularization
by analytic continuation in $d=4 - 2\ep$ space-time dimensions, and we use the
customary scheme of conventional dimensional regularization (CDR). The quantity
${\cal M}_{c{\bar c}\to Q {\bar Q}}(p_1, p_2;p_3,p_4) \equiv 
{\cal M}_{c{\bar c}\to Q {\bar Q}}(\{ p_i \})$
in the right-hand side of  Eq.~(\ref{mtil})
is the renormalized on-shell scattering amplitude 
\cite{Czakon:2008zk, Bonciani:2008az},
and it has the perturbative expansion
\begin{align}
{\cal M}_{c{\bar c}\to Q {\bar Q}}(\{ p_i \}) \!
&= \as(\mu_R^2) \,\mu_R^{2\ep} 
\left[
{\cal M}_{c{\bar c}\to Q {\bar Q}}^{\,(0)}(\{ p_i \})
\!+ \sum_{n=1}^{\infty} \left(\frac{\as(\mu_R^2)}{2\pi}\right)^{\!\!n}
\!\!{\cal M}_{c{\bar c}\to Q {\bar Q}}^{\,(n)}(\{ p_i \}; \mu_R)
\right] .
\label{ampli}
\end{align}
The perturbative expansion of $\widetilde{\cal M}_{c{\bar c}\to Q {\bar Q}}$
is completely analogous to that in Eq.~(\ref{ampli}), with 
$\widetilde{\cal M}_{c{\bar c}\to Q {\bar Q}}^{\,(0)} =
{\cal M}_{c{\bar c}\to Q {\bar Q}}^{\,(0)}$ and the replacement
${\cal M}_{c{\bar c}\to Q {\bar Q}}^{\,(n)} \to 
\widetilde{\cal M}_{c{\bar c}\to Q {\bar Q}}^{\,(n)}$ ($n\geq 1$).
Using Eq.~(\ref{mtil}), we can readily obtain 
$\widetilde{\cal M}_{c{\bar c}\to Q {\bar Q}}^{\,(n)}$ as a function of 
${\cal M}_{c{\bar c}\to Q {\bar Q}}^{\,(k)}$ and 
$\widetilde{\i}^{(k)}_{c{\bar c}\to Q {\bar Q}}$ with $k \leq n$.
For instance, at the NLO level we have
\beq
\label{mtil1}
\widetilde{\cal M}_{c{\bar c}\to Q {\bar Q}}^{\,(1)} =
{\cal M}_{c{\bar c}\to Q {\bar Q}}^{\,(1)} - 
\widetilde{\i}^{(1)}_{c{\bar c}\to Q {\bar Q}} 
\;{\cal M}_{c{\bar c}\to Q {\bar Q}}^{\,(0)} \;\;.
\eeq
The renormalized virtual amplitude ${\cal M}_{c{\bar c}\to Q {\bar Q}}$ still
has IR divergences in the form of $1/\ep$ poles.
The subtraction operator $\widetilde{\i}_{c{\bar c}\to Q {\bar Q}}(\as,\ep)$, 
which originates from real emission contributions to the \qt\ cross section, 
also contains IR divergences. More precisely, it exactly includes the IR divergent
terms that are necessary to cancel the IR divergences of the amplitude
${\cal M}_{c{\bar c}\to Q {\bar Q}}$, and it includes additional IR finite terms
that are specific of the \qt\ cross section in Eq.~(\ref{crosssec}).
Therefore, the hard-virtual amplitude 
$\widetilde{\cal M}_{c{\bar c}\to Q {\bar Q}}$ can be safely computed in the 
limit $\ep \to 0$. The expressions of
$\left( \h \,\de \right)$ in Eqs.~(\ref{Hq}) and 
(\ref{Hg}) have to be evaluated by setting $\ep =0$
in $\widetilde{\cal M}_{c{\bar c}\to Q {\bar Q}}$, although the 
four-dimensional limit $\ep \to 0$ is not explicitly denoted in the right-hand
side of those equations.
We note that the {\em all-order} factors ${\cal M}$, $\widetilde{\i}$ and, hence,
$\widetilde{\cal M}$ are renormalization-group invariant quantities
(i.e., they are independent of $\mu_R$). Their dependence on $\mu_R$ only
appears throughout the fixed-order truncation of the perturbative series in powers
of $\as(\mu_R^2)$ (see Eqs.~(\ref{itilall}), (\ref{ampli}) and (\ref{mtil1})).
We also remark that the operator $\widetilde{\i}_{c{\bar c}\to Q {\bar Q}}$
is completely independent of the spin of the four external hard partons of the
process $c{\bar c}\to Q {\bar Q}$. In particular, the gluon Lorentz
indices $\{ \mu_i^\prime, \nu_i^\prime \}$ $(i=1,2)$ of 
$\widetilde{\cal M}_{gg\to Q {\bar Q}}$ in Eq.~(\ref{Hg}) are exactly those 
of the corresponding amplitude
${\cal M}_{gg\to Q {\bar Q}}$ in the right-hand side of Eq.~(\ref{mtil}).

In the region of very small values of $q_T$, $q_T \ltap \Lambda$ ($\,\Lambda$
is the QCD scale) or, equivalently, at very large values of $b\,$ 
($\,b \Lambda \gtap 1$), the perturbative computation of the 
$q_T$ cross section has to be supplemented with non-perturbative corrections.
Non-perturbative contributions are embodied in
transverse-momentum dependent (TMD) parton densities
\cite{Collins:1981uk, Kodaira:1981nh, Catani:vd}
that can be used to express the $q_T$ cross section 
in the small-$q_T$ region through
TMD factorization
(see Ref.~\cite{Collins:2011zzd} and references therein).
In the context of TMD factorization, roughly speaking, the factor 
${\sqrt {S_c(M,b)}} \;C(\as(b_0^2/b^2)) \otimes f(b_0^2/b^2)$
(here $C$ denotes the collinear functions in Eqs.~(\ref{what}) and (\ref{whatgg}),
and the symbol `$\otimes$' denotes the convolution with respect to the momentum
fraction $z$) of the resummation formula (\ref{crosssec})
arises from the TMD parton density 
\cite{Becher:2010tm, GarciaEchevarria:2011rb}
in the region $b \Lambda \ltap 1$. In the case of production of a colourless
system $F$, the resummation formula (\ref{crosssec}) has no other $b$ dependent
factors. In the case of \qq\ production, the presence in Eq.~(\ref{crosssec})
of one additional $b$ dependent factor, the soft-parton factor $\de$,
is consistent with a breakdown (in weak form) 
\cite{Collins:2007nk}
of TMD factorization. In the production processes of strongly interacting 
systems
(such as \qq\ pairs), TMD parton densities have to be supplemented with
additional and process dependent non-perturbative factors 
\cite{Bomhof:2004aw}. 
As we have previously discussed, the breakdown (in strong form) 
\cite{Rogers:2010dm}
of TMD factorization can have connection with
high-order structures in 
transverse-momentum resummation. 

In the framework of TMD factorization, azimuthal correlations in heavy-quark
production processes at small $q_T$ have been discussed in 
Ref.~\cite{Boer:2010zf}. The azimuthal dependence that is explicitly worked out
in Ref.~\cite{Boer:2010zf} arises from TMD factorization and, therefore, it is
consistent with the azimuthal dependence (see Ref.~\cite{Catani:2010pd} and 
the discussion below Eq.~(\ref{dbten})) driven by the gluon collinear function 
$C_{g \, a}^{\mu \, \nu}$ of the resummation formula (\ref{crosssec}).
The complete structure of azimuthal correlations in Eq.~(\ref{crosssec})
receives additional contributions from the soft-parton factor
$\de$ (see Eqs.~(\ref{phiq})--(\ref{phig}) and the accompanying discussion).
Since $\de$ is related to TMD factorization breaking effects, these (colour
charge dependent) azimuthal correlations cannot originate from
process-independent TMD parton densities (see also Sect.~V of the second paper 
in Ref.~\cite{Boer:2010zf}).

\section{Explicit results for the resummation coefficients}

In this Section we present our explicit analytic results for the resummed cross section in
Eq.~(\ref{crosssec}) up to NLO and NNLL accuracy. To this purpose we can exploit
the knowledge of the universal (process-independent) factors $S_c, C_{c\,a}$ and
$C_{g\,a}^{\mu \nu}$ up to NNLL+NNLO. The Sudakov form factor $S_c(M,b)$ has an
all-order representation \cite{Collins:1981uk}
(see, e.g., Eq.~(8) in Ref.~\cite{Catani:2013tia})
that is fully specified by two perturbative functions $A_c(\as)$ and $B_c(\as)$.
The corresponding perturbative coefficients 
$A^{(1)}_c, B^{(1)}_c, A^{(2)}_c$
\cite{Kodaira:1981nh, Catani:vd},  $B^{(2)}_c$ 
\cite{Davies:1984hs, deFlorian:2000pr} 
and $A^{(3)}_c$ \cite{Becher:2010tm} are explicitly known, and they determine 
$S_c(M,b)$ up to NNLL accuracy.
The partonic collinear functions 
$C_{c\,a}$ $(c=q,{\bar q})$ and
$C_{g\,a}^{\mu \nu}$ in Eqs.~(\ref{what}) and (\ref{whatgg})
are known 
\cite{Catani:2007vq, Catani:2009sm, Catani:2011kr, Gehrmann:2012ze}
up to ${\cal O}(\as^2)$ (i.e., NNLO).
The two computations in Refs.~\cite{Catani:2011kr} and \cite{Gehrmann:2012ze}
are fully independent and they lead to results in full agreement.
As we have already recalled, the determination of the individual (separate) 
factors of the resummation formula (\ref{crosssec}) requires the specification of
a resummation scheme 
\cite{Catani:2000vq}.
The collinear functions of Ref.~\cite{Gehrmann:2012ze}, which refer to
transverse-momentum resummation according to the formulation of 
Ref.~\cite{Becher:2010tm}, are eventually related to our functions 
$C_{c\,a}$ and $C_{g\,a}^{\mu \nu}$ \cite{Catani:2011kr}
throughout a transformation of resummation scheme. In the following, to present 
our results, we explicitly consider the `hard scheme' used in 
Ref.~\cite{Catani:2013tia}. The expressions of the universal factors 
$S_c, C_{c\,a}$ and $C_{g\,a}^{\mu \nu}$ in the hard scheme can be found in 
Ref.~\cite{Catani:2013tia}. The remaining perturbative ingredients of the \qq\
resummation formula (\ref{crosssec}) are the hard factor $\h$
(i.e, the subtraction operator $\widetilde{\i}_{c{\bar c}\to Q {\bar Q}}$),
the soft evolution factor $\v$ (i.e., the soft anomalous dimension $\g$)
and the soft azimuthal-correlation factor $\dcor$.
We have computed $\widetilde{\i}_{c{\bar c}\to Q {\bar Q}}$, $\g$ and $\dcor$
at ${\cal O}(\as)$, and we have determined $\g$ at ${\cal O}(\as^2)$ by 
relating it to
the ${\cal O}(\as^2)$ computation 
\cite{Mitov:2009sv, Ferroglia:2009ep, Ferroglia:2009ii}
of the IR anomalous dimension of the
scattering amplitude ${\cal M}_{c{\bar c}\to Q {\bar Q}}$:
these results complete the
evaluation of the \qq\ resummation formula (\ref{crosssec}) up to NNLL+NLO.
Using the hard scheme, the 
results of our computation are presented
below (see Eqs.~(\ref{i1til}), (\ref{ga1}), (\ref{d1cor}) and (\ref{gtot})).

The colour operators $\widetilde{\i}_{c{\bar c}\to Q {\bar Q}}$, $\g$ and 
$\dcor$ depend on the colour charges $(\T_i)^a$ ($a=1,\dots,N_c^2-1$ is the
colour index of the radiated gluon) of the four $(i \leq 4)$ radiating partons
$c,{\bar c},Q,{\bar Q}$. Using the colour space formalism of
Ref.~\cite{Catani:1996vz}, the colour charge $(\T_i)^a$ is a colour matrix in
either the fundamental (if $i$ is a quark) or adjoint (if $i$ is a gluon)
representation of $SU(N_c)$ in QCD with $N_c$ colours. Note that the colour flow
of the process $c{\bar c}\to Q {\bar Q}$ is treated as `outgoing', so that 
$\T_3$ and $\T_4$ are the colour charges of $Q(p_3)$ and  ${\bar Q}(p_4)$, while
$\T_1$ and $\T_2$ are the colour charges of the anti-partons
${\bar c}(-p_1)$ and $c(-p_2)$ in Eq.~(\ref{partpro}). According to this 
notation, colour conservation implies $\sum_{i=1}^4 \T_i \;| \dots\rangle = 0$,
where $| \dots\rangle$ is a colour-singlet state vector, such as
$| {\cal M}_{c{\bar c}\to Q {\bar Q}} \rangle$ 
or $| \widetilde{\cal M}_{c{\bar c}\to Q {\bar Q}} \rangle$. We also define
$\T_i \cdot \T_j \equiv (\T_i)^a (\T_j)^a$ and, in particular, $\T_i^{\,2}$
is a $c$-number term (more precisely, $\T_i^{\,2}$ is a multiple of the unit
matrix in colour space) given by the Casimir factor ($C_F$ or $C_A$) of the
corresponding representation of $SU(N_c)$. We have $\T_1^{\,2}=\T_2^{\,2}=C_F=
(N_c^2 -1)/(2N_c)$ in the $q{\bar q}$ annihilation channel, 
$\T_1^{\,2}=\T_2^{\,2}=C_A=N_c$ in the gluon fusion channel, whereas 
$\T_3^{\,2}=\T_4^{\,2}=C_F$.
Considering the kinematics of the process $c{\bar c}\to Q {\bar Q}$ in
Eq.~(\ref{partpro}), four-momentum conservation leads to the relations
$\; y_3 -y = y -y_4= y_{34}/2$, 
$\;{\bf p}_{{\rm \bf T}3}^2 = {\bf p}_{{\rm \bf T}4}^2 \equiv \pts$
and the heavy-quark transverse mass $m_T= \sqrt{m^2+\pts}$ is related to 
$y_{34}$ by using $M=2m_T \cosh(y_{34}/2)$.
Using these kinematical relations the operators
$\widetilde{\i}_{c{\bar c}\to Q {\bar Q}}$, $\g$ and 
$\dcor$ can eventually be expressed in term of the two independent variables
$y_{34}$ and $2m/M$ (or, equivalently, the relative velocity $v$ in 
Eq.~(\ref{relv})). As already discussed, $\dcor$ additionally depends on the
relative azimuthal angle $\phi_{3b}$ (or, equivalently, $\phi_{4b}$).

The first-order term $\widetilde{\i}^{(1)}$ of the subtraction operator
$\widetilde{\i}_{c{\bar c}\to Q {\bar Q}}$ in Eqs.~(\ref{mtil}) and 
(\ref{itilall}) has the following form:
\begin{align}
\label{i1til}
\widetilde{\i}^{(1)}_{c{\bar c}\to 
Q {\bar Q}}\left(\ep,\frac{M^2}{\mu_R^2}\right) 
= - \frac{1}{2}
\left( \frac{M^2}{\mu_R^2}\right)^{\!\!-\ep} \!\left\{
\left(\frac{1}{\ep^2} +i\pi \frac{1}{\ep} 
-\frac{\pi^2}{12}\right) (\T_1^2 +\T_2^2)
+ \frac{2}{\ep} \,\gamma_c  
- \frac{4}{\ep} \;\g^{(1)}(y_{34}) + \F(y_{34}) \right\} .
\end{align} 
The flavour dependent coefficients $\gamma_c$ ($c=q,{\bar q},g$)
originate from collinear radiation:
the explicit values of these coefficients 
are $\gamma_q=\gamma_{\bar q}=3C_F/2$
and $\gamma_g= (11C_A-2N_f)/6$, and $N_f$ is the number of flavours of 
massless quarks (e.g., $N_f=5$ in the case of $t{\bar t}$ production).
The IR finite contribution $\F$ to Eq.~(\ref{i1til}) is
\begin{align}
\label{F1T}
\F(y_{34}) = (\T_3^2 + \T_4^2)
\;\ln\left(\frac{m_T^2 }{m^2}\right)
+ \;(\T_3 + \T_4)^2 \;
{\rm Li}_2\!\left(-\frac{\pts}{m^2} \right)
+ \T_3 \cdot \T_4 \;\frac{1}{v} \,L_{34} \;\;,
\end{align}
where the function $L_{34}$ is
\begin{align}
\label{L34}
L_{34}&=\ln\left( \frac{1+v}{1-v} \right)
\, \ln \left(\frac{m_T^2}{m^2}\right)
- 2 \,{\rm Li}_2\left( \frac{2 v}{1+v}\right)
- \frac{1}{4}\ln^2\left( \frac{1+v}{1-v} \right) \nn\\
&+2\left[ \,{\rm Li}_2\left( 1 - 
\sqrt{\frac{1-v}{1+v}}\, e^{\,y_{34}} \right)
+ \,{\rm Li}_2\left( 1 - 
\sqrt{\frac{1-v}{1+v}}\,e^{-y_{34}} \right)
+ \frac{1}{2}\,y_{34}^2\right]  
\end{align}
and ${\rm Li}_2$ is the customary dilogarithm function,
${\rm Li}_2(z) = -\int_0^z \frac{dt}{t} \,\ln(1-t)$.

The colour operator  $\g^{(1)}(y_{34})$ in the right-hand side of 
Eq.~(\ref{i1til}) is exactly equal to the first-order term of the soft anomalous
dimension in Eq.~(\ref{gaeq}), and its explicit form is
\begin{align}
\g^{(1)}(y_{34}) = -\f{1}{4}& 
\left\{  (\T_3^2 + \T_4^2)\;(1 - i \pi)
+ \sum_{\substack{i=1,2 \\ j =3,4}} 
\;\T_i \cdot \T_j \,\ln\frac{(2p_i\cdot p_j)^2}{M^2 m^2} \right. \nn\\
&+ \left. 
2\;\T_3 \cdot \T_4 \left[ \frac{1}{2v} \ln\left(\frac{1+v}{1-v}\right) - 
i \pi \left( \frac{1}{v} + 1 \right)
\right]
\right\} \;\;.
\label{ga1}
\end{align}
We note that the second term in the right-hand side of Eq.~(\ref{ga1})
can be rewritten as
\beq
\sum_{\substack{i=1,2 \\ j =3,4}} 
\;\T_i \cdot \T_j \,\ln\frac{(2p_i\cdot p_j)^2}{M^2 m^2} = 
(\T_3 + \T_4)^2 \,\ln\left(\frac{m_T^2 }{m^2}\right)
- (\T_1 - \T_2) \cdot (\T_3 - \T_4) \;y_{34} \;\;,
\eeq
where we have simply used colour conservation and kinematical relations.

The expression of $\widetilde{\i}^{(1)}_{c{\bar c}\to Q {\bar Q}}$
in Eq.~(\ref{i1til}) contains IR divergent terms in the form of
double and single poles $1/\ep^2$ and $1/\ep$. We have explicitly
checked that these IR
divergent terms are exactly those that control the factorized IR structure 
\cite{Catani:2000ef}
of general one-loop scattering amplitudes with massive external partons. 
This directly
proves that the one-loop hard-virtual amplitude 
$\widetilde{\cal M}_{c{\bar c}\to Q {\bar Q}}^{\,(1)}$ in Eq.~(\ref{mtil1})
is IR finite in the limit $\ep \to 0$.
The right-hand side of Eq.~(\ref{i1til}) also contains IR finite contributions.
As previously discussed (see, e.g., the first paragraph of this Section),
these IR finite contributions depend on the specification of the resummation
scheme. The explicit expression in the right-hand side of Eq.~(\ref{i1til})
is specific of the hard scheme \cite{Catani:2013tia}, 
supplemented with the property in Eq.~(\ref{dav}). Since 
$\widetilde{\i}^{(1)}_{c{\bar c}\to Q {\bar Q}}$ does not depend on $\bf b$,
this scheme choice uniquely determines how IR finite contributions
are split between $\widetilde{\i}^{(1)}_{c{\bar c}\to Q {\bar Q}}$
and $\dcor^{(1)}$.

The soft-parton operator $\dcor$ in Eq.~(\ref{deq}) also depends on the
relative azimuthal angle $\phi_{3b}$ (or, equivalently, $\phi_{4b}$).
The expression of the first-order term $\dcor^{(1)}$ 
is quite involved.
To shorten the notation we define the auxiliary variable $\c3b$,
\beq
\c3b = \frac{\sqrt {\pts}}{m} \;\cos(\phi_{3b}) = 
- \,\frac{\sqrt {\pts}}{m} \;\cos(\phi_{4b})\;\;.
\eeq
We obtain the following result:
\begin{align}
\dcor^{(1)}(\phi_{3b}, y_{34}) &= (\T_3^2 + \T_4^2) \left[
\frac{\c3b \,{\rm arcsinh}\left(\c3b\right)}{\sqrt {1 + \c3b^2}}\;
- \frac{1}{2} \ln\left(\frac{m_T^2}{m^2}\right)
\right]
\nn \\
&- (\T_3 + \T_4)^2  
\left(
{\rm arcsinh}^2\left(\c3b\right)
+ \frac{1}{2} {\rm Li}_2\!\left(-\frac{\pts}{m^2}\right)
\right)+ \frac{1}{2v}\,\T_3 \cdot \T_4 \,\left(\L34phi-L_{34}\right)\;\;,
\label{d1cor}
\end{align}
where $L_{34}$ is given in Eq.~(\ref{L34}). The function $\L34phi$ is
\begin{equation}
\label{L34phi}
\L34phi={\rm Sign}(c_{3b})\Big[L_\xi \left(\xi(c_{3b},\alpha_{34}),\alpha_{34}\right)
-L_\xi\left(\xi(-c_{3b},\alpha_{34}),\alpha_{34}\right)\Big]
\end{equation}
with
\begin{equation}
L_\xi(\xi,\alpha)=\f{1}{2}\ln^2\f{\xi(1+\xi)}{\alpha+\xi}-\ln^2 \f{\xi}{\alpha+\xi}
+2\left[{\rm Li}_2(-\xi)-{\rm Li}_2\left(\f{\alpha+\xi}{\alpha-1}\right)+\ln(\alpha+\xi)\ln(1-\alpha)\right]
\end{equation}
and
\begin{equation}
\xi(c,\alpha)=\left(c+\sqrt{1+c^2}\right)\left(c+\sqrt{\alpha+c^2}\right)\;\;\;\;\;,
\;\;\;\;\;
\alpha_{34}=\f{2\,\sqrt{1-v^2}}{1-\sqrt{1-v^2}}\, c_{3b}^2\;\; .
\end{equation}
By simple inspection of Eq.~(\ref{d1cor}), we can observe that the azimuthal
dependence of $\dcor^{(1)}$ is quite complex and entangled with the colour
correlation factor $\T_3 \cdot \T_4$: this is a consequence of its dynamical
origin from the specific angular pattern of soft-gluon radiation 
in \qq\ production. 
We note that the expression in Eq.~(\ref{d1cor}) has a vanishing azimuthal
average (i.e., $\langle \;\dcor^{(1)}(\phi_{3b}, y_{34}) \rangle_{{\rm av.}}
= 0$) and, therefore, the property in Eq.~(\ref{dav}) is fulfilled.

The first-order term $\g^{(1)}$ (see Eq.~(\ref{ga1})) of the soft anomalous
dimension controls (through Eq.~(\ref{veq})) $q_T$ resummation up to NLL
accuracy. The second-order term $\g^{(2)}$ of the soft anomalous dimension
in Eq.~(\ref{gaeq}) is also necessary to determine the NNLL contributions.
Both $\g^{(1)}$ and $\g^{(2)}$ are related to the IR singularities of the
virtual scattering amplitude ${\cal M}_{c{\bar c}\to Q {\bar Q}}$, which are
explicitly known at one-loop \cite{Catani:2000ef} and two-loop \cite{Mitov:2009sv, Ferroglia:2009ep, Ferroglia:2009ii} order:
exploiting this knowledge, we have determined $\g^{(2)}$.
We obtain the result
\begin{equation}
\label{gtot}
\g(\as;y_{34}) = \frac{1}{2} 
\;{\bf \Gamma}_{c{\bar c}\to Q {\bar Q}}^{\,\rm sub.}(\as;y_{34}) - 
\left(\frac{\as}{\pi}\right)^2 \;\frac{1}{4}
\left( \; \left[ \g^{(1)}(y_{34}) \,, \F(y_{34}) \right] + 
\pi \beta_0 \F(y_{34})
\right) +
{\cal O}(\as^3) \;,
\end{equation}
where $12 \pi \beta_0 = 11 N_c -2N_f$, and $\F$ and $\g^{(1)}$ are given 
in Eqs.~(\ref{F1T}) and (\ref{ga1}). The `subtracted' anomalous dimension
${\bf \Gamma}_{c{\bar c}\to Q {\bar Q}}^{\,\rm sub.}$
is directly related to the IR anomalous dimension of 
Ref.~\cite{Ferroglia:2009ii} (as explained below). The perturbative expansion 
of the right-hand side of Eq.~(\ref{gtot}) includes both the first-order and
second-order terms $\g^{(1)}$ and $\g^{(2)}$ (obviously,
${\bf \Gamma}_{c{\bar c}\to Q {\bar Q}}^{\,\rm sub.}= 2 (\as/\pi) \g^{(1)}
+ {\cal O}(\as^2)$), while terms at ${\cal O}(\as^3)$ and beyond are neglected.

As we have previously discussed, the evolution operator $\v$ and, thus,
$\g$ are essentially determined by virtual soft-parton radiation through the
cancellation mechanism of the IR singularities of the scattering amplitude
${\cal M}_{c{\bar c}\to Q {\bar Q}}$. This origin in 
manifest in Eq.~(\ref{i1til}), where $\g^{(1)}$ enters as coefficient of the
single pole $1/\ep$. In particular, setting $\g^{(1)}=0$ in Eq.~(\ref{i1til}),
the IR divergences of
the subtraction operator $\widetilde{\i}^{(1)}_{c{\bar c}\to Q {\bar Q}}$
would be exactly equal to those of the analogous subtraction operator
\cite{Catani:2013tia}
for the production of a colourless system $F$.
This means that $\g^{(1)}$ controls the IR divergences due to soft wide-angle
radiation in the process $c{\bar c}\to Q {\bar Q}$. This origin of $\g$
remains valid at higher perturbative orders, and it leads to the contribution 
${\bf \Gamma}_{c{\bar c}\to Q {\bar Q}}^{\,\rm sub.}$ in Eq.~(\ref{gtot}).
The subtracted anomalous dimension 
${\bf \Gamma}_{c{\bar c}\to Q {\bar Q}}^{\,\rm sub.}$ is given by the following
relation:
\begin{equation}
\label{gasub}
{\bf \Gamma}_{c{\bar c}\to Q {\bar Q}}^{\,\rm sub.}(\as;y_{34}) =
{\bf \Gamma}(\mu) -  
\left[ \frac{1}{2} (\T_1^2 + \T_2^2) \,\gamma_{\rm cusp}(\as) 
\left( \ln \frac{M^2}{\mu^2} - i \pi \right) + 2 \gamma^c(\as) \right] \;\;,
\end{equation}
where the terms on the right-hand side are written by exactly using the 
notation of Eq.~(5) of Ref.~\cite{Ferroglia:2009ii}.
The term ${\bf \Gamma}(\mu)$ is the anomalous-dimension matrix that controls the
IR divergences of the scattering amplitude 
${\cal M}_{c{\bar c}\to Q {\bar Q}}$, while the square-bracket term on the 
right-hand side of Eq.~(\ref{gasub}) is the corresponding expression of 
${\bf \Gamma}(\mu)$ for a generic process $c{\bar c}\to F$
(where the system $F$ is colorless). The square-bracket term is 
the contribution of soft and collinear radiation from the colliding partons
$c$ and ${\bar c}$. In Eq.~(\ref{gasub}), this contribution is subtracted from
${\bf \Gamma}(\mu)$, so that 
${\bf \Gamma}_{c{\bar c}\to Q {\bar Q}}^{\,\rm sub.}$ embodies the remaining IR
effects due to soft wide-angle radiation in the process 
$c{\bar c}\to Q {\bar Q}$. We note that the subtraction in Eq.~(\ref{gasub})
exactly corresponds to the splitting procedure used in Eq.~(57) of 
Ref.~\cite{Li:2013mia} to introduce the anomalous dimension 
${\bf \gamma}_{i \bar i}^h$: therefore, we have 
${\bf \Gamma}_{c{\bar c}\to Q {\bar Q}}^{\,\rm sub.} = 
{\bf \gamma}_{c \bar c}^h$. The expression of ${\bf \Gamma}(\mu)$ at 
${\cal O}(\as^2)$ is computed and explicitly given in 
Ref.~\cite{Ferroglia:2009ii}. This expression (which is too long to be reported
here) straightforwardly leads to the ${\cal O}(\as^2)$ term of 
${\bf \Gamma}_{c{\bar c}\to Q {\bar Q}}^{\,\rm sub.}$ in Eq.~(\ref{gasub})
and to the ensuing contribution in Eq.~(\ref{gtot}). The additional contribution
to $\g$ in Eq.~(\ref{gtot}) is proportional to $\F$, and it is due to the
corresponding IR finite contribution to 
$\widetilde{\i}^{(1)}_{c{\bar c}\to Q {\bar Q}}$ in Eq.~(\ref{i1til}).
Both contributions eventually originate from the property in Eq.~(\ref{dav})
of the soft-parton factor $\dcor$.

We note that the first-order term $\g^{(1)}$ of the soft anomalous
dimension includes (see Eq.~(\ref{ga1})) an absorptive (antihermitian) term
of the type ${\bf \Gamma}^{(1)}_{(C)} \propto i \;\T_3 \cdot \T_4$ (it is due
to the non-abelian QCD analogue of the QED Coulomb phase)
that involves colour correlations between two partons. Owing to its
antihermitian character, ${\bf \Gamma}^{(1)}_{(C)}$ gives a vanishing
contribution 
(see the factors in Eqs.~(\ref{Hq}), (\ref{Hg}) and (\ref{deeq}))
to the \qq\ cross section at the NLO. Nonetheless, 
${\bf \Gamma}^{(1)}_{(C)}$ does contribute to the singular component of the
$q_T$ cross section at higher perturbative orders.
A related comment applies to a term, ${\bf \Gamma}^{(2)}_{(3)} \propto f^{a b c}
 \;\T_1^a \, \T_3^b \, \T_4^c$, that contributes to the second-order anomalous
dimension $\g^{(2)}$. The triple colour correlation term 
${\bf \Gamma}^{(2)}_{(3)}$ originates from the commutator 
$\left[ \g^{(1)} \,, \F \right]$ in the right-hand side of
Eq.~(\ref{gtot}) and from a corresponding term in 
${\bf \Gamma}_{c{\bar c}\to Q {\bar Q}}^{\,\rm sub.}$
(see Eq.~(\ref{gasub}) and the expression of ${\bf \Gamma}(\mu)$ in Eq.~(5)
of Ref.~\cite{Ferroglia:2009ii}). In the computation of the \qq\ cross section
(see the factors in Eqs.~(\ref{Hq}), (\ref{Hg}) and (\ref{deeq})),
${\bf \Gamma}^{(2)}_{(3)}$ gives a vanishing contribution at the NNLO
\cite{Czakon:2009zw, Ferroglia:2009ii, Czakon:2013hxa}.
This follows from the fact that the tree-level amplitude 
${\cal M}_{c{\bar c}\to Q {\bar Q}}^{\,(0)} = 
\widetilde{\cal M}_{c{\bar c}\to Q {\bar Q}}^{\,(0)}$ is real and, therefore,
$\langle \widetilde{\cal M}_{c{\bar c}\to Q {\bar Q}}^{\,(0)} | \,
{\bf \Gamma}^{(2)}_{(3)} \, 
| \widetilde{\cal M}_{c{\bar c}\to Q {\bar Q}}^{\,(0)} \rangle = 0$
\cite{Forshaw:2012bi, Czakon:2013hxa}.
At higher perturbative orders the resummation factors in 
Eqs.~(\ref{Hq}), (\ref{Hg}) and (\ref{deeq}) include additional absorptive
terms (e.g., the one-loop amplitude 
$\widetilde{\cal M}_{c{\bar c}\to Q {\bar Q}}^{\,(1)}$ is not purely real) and,
therefore, ${\bf \Gamma}^{(2)}_{(3)}$ gives non-vanishing contributions to the
$q_T$ cross section beyond the NNLO level (see the related discussion in the 
Note Added of Ref.~\cite{Catani:2011st}).

Transverse-momentum resummation for \qq\ production has been studied in 
Refs.~\cite{Zhu:2012ts, Li:2013mia}. The framework developed in  
Refs.~\cite{Zhu:2012ts, Li:2013mia} is an extension of the SCET formulation
of \qt\ resummation that was presented in 
Ref.~\cite{Becher:2010tm} for the cases of DY and Higgs boson production.
The authors of Refs.~\cite{Zhu:2012ts, Li:2013mia} consider the
azimuthally-averaged \qt\ cross section and present results at the NLO and NNLL
accuracy. We have performed a comparison between those results and our results,
and we find full agreement. The comparison poses no difficulties since, as we
have discussed, we can straightforwardly obtain the 
azimuthally-averaged \qt\ cross section by integrating the resummation formula
(\ref{crosssec}). In particular, at NLO and NNLL accuracy, we can simply set
$\de = \v^\dagger \, \v$ (i.e., $\dcor=1$) in Eq.~(\ref{crosssec})
(this follows from Eq.~(\ref{aveq}) and from the fact the inequality in
Eq.~(\ref{aveg}) is due to terms of ${\cal O}(\as^2)$, which start to contribute
at the NNLO and beyond NNLL accuracy). We note that the various 
(hard, soft, collinear) resummation
factors in our Eq.~(\ref{crosssec}) and those in Ref.~\cite{Li:2013mia} 
are separately different, since they correspond to the use of different
resummation schemes.

As discussed in Sect.~\ref{sec:res}, the results presented in this Section 
are obtained by using soft/collinear factorization formulae
\cite{Campbell:1997hg, Bern:1999ry, Kosower:1999rx, Catani:1999ss, 
Catani:2000pi, Czakon:2011ve}, and they can be extended to the complete NNLO
level through the evaluation of the ${\cal O}(\as^2)$-terms 
$\dcor^{(2)}$ and $\widetilde{\i}^{(2)}$ in 
Eqs.~(\ref{deq}) and (\ref{itilall}). The extension, which does not require
further conceptual steps, is certainly complex from the computational
viewpoint.

\section{Summary}

In this paper we have considered the transverse-momentum distribution of a
heavy-quark pair produced
in hadronic collisions.
As in the case of simpler processes, such as the hadroproduction of a system of
non-strongly interacting particles,
the perturbative QCD computation of the
$q_T$ cross section is affected by large logarithmic terms that need be
resummed to all perturbative orders.
We have discussed the new issues that arise in the case of heavy-quark 
production, and we have 
presented our all-order resummation formula (see Eq.~(\ref{crosssec})) for
the logarithmically-enhanced contributions.
The main differences with respect to the production of colourless 
systems
is the appearance of the soft factor $\de$ (see Eq.~(\ref{deeq}))
that is due to soft-parton radiation at large angles with respect to the
direction of the colliding hadrons (partons).
The factor $\de$ embodies the effect of 
soft radiation from the heavy-quark final state and from initial-state 
and final-state
interferences. The dynamics of soft-parton radiation produces colour-dependent
azimuthal correlations in the small-$q_T$ region. This azimuthal dependence is
fully taken into account by the resummation formula and it is embodied in the soft-parton
factor $\de$: the dependence is 
controlled by the colour operator $\dcor$
and it is factorized with respect to the
colour (soft) evolution factor $\v$ (see Eq.~(\ref{deeq})).
We have shown how the azimuthal correlations of soft-parton origin are 
entangled with the azimuthal dependence due
to gluonic collinear radiation (see Eq.~(\ref{phig})),
and we have discussed the ensuing effect on the azimuthally-averaged
$q_T$ cross section.
We have presented
the explicit results of the perturbative coefficients of the resummation formula
up to NLO and NNLL accuracy
(see Eqs.~(\ref{i1til}), (\ref{ga1}), (\ref{d1cor}) and (\ref{gtot})).

Transverse-momentum resummation for heavy-quark production is important
for phenomenological applications through resummed calculations 
\cite{Li:2013mia},
especially for the production of top-quark pairs.
Given the huge amount of
top-quark pairs that have been produced at the LHC in its first run, and the
even higher number of $t{\bar t}$ events that are expected at
$\sqrt{s}=13~(14)$ TeV, the possibility of relying on accurate computations of
the transverse-momentum spectrum of the $t{\bar t}$ pair down to the low-$q_T$
region 
is very relevant 
for physics studies within and beyond the SM.
                                                                                
We point out that the $q_T$ resummation formalism for \qq\ production
has
implications not only for resummed calculations but also for fixed-order
computations up to NNLO. The $q_T$ subtraction formalism \cite{Catani:2007vq}
is an efficient method to perform fully-exclusive NNLO computations 
of hard-scattering processes, and it is
based on the knowledge of the small-$q_T$ limit of the 
transverse-momentum cross section of the corresponding process. 
In the case of the production of colourless systems,
thanks to the complete understanding of the all-order structure of the
large logarithmic terms,
the method is fully developed up to NNLO.
The resummation formula presented in Eq.~(\ref{crosssec}) makes possible to
apply the $q_T$ subtraction formalism also to heavy-quark production at NNLO, 
once the explicit results of the resummation factors at the 
corresponding order will be available.

\noindent {\bf Acknowledgements}.
This research is supported in part by the Swiss National Science Foundation (SNF) under contract 200021-144352 and by 
the Research Executive Agency (REA) of the European Union under the Grant Agreement number PITN-GA-2010-264564 ({\it LHCPhenoNet}).

\end{document}